\documentclass[11pt]{article}
\usepackage[utf8]{inputenc}

\usepackage[margin=1in]{geometry}

\usepackage{amsmath, amssymb}

\usepackage{enumitem}
\usepackage{tabularx}
\usepackage{graphicx}
\graphicspath{{./}{figures/}{../}{../figures/}}

\usepackage[colorlinks=true, allcolors=blue]{hyperref}

\usepackage{caption}
\usepackage{subcaption}
\usepackage{comment}

\usepackage[style=ext-numeric, giveninits=true, sorting=none, sortcites=true, articlein=false]{biblatex}
\usepackage{xcolor}
\usepackage{newunicodechar}
\newunicodechar{́}{\'{}}

\newcommand{\der}[0]{\,\mathrm{d}}

\title{Effect of environmental variation on the benefits of learning}
\author{Xiao Zhou, BingKan Xue}
\date{Department of Physics, University of Florida, Gainesville, FL 32611, USA}

\begin{document}

\maketitle

\begin{abstract}
Signal recognition plays a critical role in species interactions and can be enhanced by learning signal characteristics through experience. In brood parasitism, host species may use visual cues to recognize and reject parasite eggs from their nests; because egg appearances vary within and between host individuals, a host can improve recognition by learning a tailored template of its own eggs. Nevertheless, constitutive and induced costs of learning may inhibit an extended learning period. We use a simple model of signal detection and learning to study how the benefits of learning are affected by different sources of variation in the learning signal. We find that phenotypic variation in egg appearances within a host hinders learning by adding noise to the signals, whereas genotypic variation between individuals promotes learning by carrying more information in the signals. Moreover, we consider environmental variation that can cause egg appearances to fluctuate across clutches over time. We find that such environmental variation reduces the fitness of learning hosts by distorting the signals, creating an effective cost that can offset the benefits of learning. Our results imply that learning or even a brief period of imprinting may be evolutionarily disfavored in homogeneous populations and variable environments.
\end{abstract}

\section{Introduction}

The ability to detect and distinguish signals is important for species' survival and reproduction in nature. For instance, predators often face the challenge of distinguishing edible prey from other species that may be toxic or unpalatable \cite{ruxton2019avoiding}. However, the availability of different prey species may vary between local communities \cite{lichtenberg2020noisy}, and phenotypic variation within prey species can further complicate their identification \cite{kikuchi2019signal}. In such circumstances, individuals may refine their decision-making by integrating information from previous experiences and using it to adjust future responses. Such learning would allow individuals to fine-tune their recognition of signals and minimize the costs of making wrong decisions \cite{shettleworth2009cognition, dukas2004evolutionary}.

The benefits of learning must be weighed against different types of potential costs associated with learning \cite{liefting2019costs, johnston1982selective}. Traditionally, constitutive costs refer to expenses required for simply having the capacity to learn, whether or not learning is actively engaged \cite{liefting2019costs, burns2011costs}. These costs may reflect investments in the learning machinery, such as neural tissues and cognitive architecture \cite{liefting2019costs, snell2011reproductive}, which can trade off against investments in other organs or reproductive capacity \cite{kotrschal2013artificial}. On the other hand, induced costs are those incurred when information is actively acquired, stored, or recalled \cite{liefting2019costs, johnston1982selective, burns2011costs}. These are typically short-term energetic expenses, such as efforts required to attend to and process new cues \cite{burns2011costs, jaumann2013energetic}, which can result in temporary declines in other physiological functions \cite{placcais2013favor}. Other types of costs have also been considered, such as sampling costs in the context of mimicry, where a predator may taste a prey that turns out to be unprofitable \cite{kikuchi2015costs}, and opportunity costs associated with suboptimal choices made during the learning process, often studied in the context of foraging \cite{stephens1991change, dunlap2016reliability}.

As a result of the trade-off between the benefits and costs of learning, different adaptation strategies may evolve. One alternative strategy is to forgo learning and rely on decision rules that are innately determined \cite{rothstein1974mechanisms, dunlap2009components}. Another strategy, called imprinting, is when learning occurs only at an early stage of life \cite{bateson1979sensitive, rothstein1978mechanisms}. For example, zebra finches memorize tutor songs during a limited juvenile period \cite{Gobes2019}, and female cichlids imprint on their mother's phenotype to determine mating preferences \cite{verzijden2007early}. As previously studied, which of the strategies is most advantageous depends on various factors, such as perception constraints \cite{Lotem_1995}, template errors \cite{shizuka2020learn}, or exploration intensity \cite{eliassen2007exploration, arehart2023minimal}.

Here, we study how the benefits of learning are affected by different sources of variation in the signals. Typically, these signals are some characteristics of other organisms that a species interacts with, such as morphological traits of prey species. These traits can vary between individuals due to genetic variation or phenotypic stochasticity \cite{Lotem_1995, lahti2002precise}. Moreover, we consider environmental variation that can influence the signals and affect the outcome of learning \cite{dunlap2009components, eliassen2007exploration, arehart2023minimal}. For example, bumblebees forage in environments where floral rewards change rapidly, so learned associations can quickly lose reliability \cite{evans2014foraging}. In such circumstances, learning can be costly because previous knowledge may quickly become outdated or misleading \cite{dunlap2016reliability}. This kind of cost is different from traditional costs of learning because it is due to uncertainty and information loss rather than energy or material expenditures.

We use brood parasitism as a model system where different learning strategies have been documented \cite{dillenseger2024active, lotem1993learning, Soler_2013}. Brood parasitism appears widely in birds, insects, and fish \cite{roldan2011parental, stevens2013bird, sato1986brood, kilner2011cuckoos}, among which avian brood parasitism has been studied the most \cite{stevens2013bird, davies2000cuckoos}. For example, common cuckoos are known to parasitize other birds' nests by removing host eggs and replacing them with their own \cite{yom1980intraspecific}. Parasites decrease the fitness of hosts not only by diverting parental effort, but many parasitic hatchlings also evict and kill the hosts' offspring \cite{kilner2005evolution, davies1988cuckoos}. In defense, hosts have evolved ways to recognize parasite eggs and reject them from the nest, using a variety of cues including color, pattern, size, and olfactory signals \cite{Rothstein1982, soler2014recognizing, fulmer2022review, spottiswoode2010visual}. Previous studies have found that host birds have innate templates to help differentiate their own eggs from parasite eggs \cite{dillenseger2024active, Rothstein1975}. Some host birds can learn from the eggs in their nests to improve recognition \cite{rothstein1974mechanisms, lotem1993learning}, and there is evidence that some host birds imprint on their first brood \cite{rothstein1978mechanisms, Lotem_1995}.

Simple models of brood parasitism and imprinting considered host and parasite eggs as having distinct appearances \cite{shizuka2020learn, lotem1993learning}, neglecting variations within each group. We consider the case where egg characteristics vary within a clutch as well as between host individuals, and their distributions overlap between the host and parasite populations \cite{Lotem_1995, davies1996recognition}. Moreover, the appearance of eggs can be influenced by environmental factors through the production of egg pigments \cite{Solomon_1991}. For example, temperature and rainfall can affect egg coloration in reed warblers \cite{aviles2007environmental}, and colder temperatures are associated with more eggshell spotting in great tits \cite{hargitai2016effects}. We will study how those different sources of variation affect the hosts' ability to recognize parasite eggs.

Learning can help host individuals develop a tailored template of their own eggs to better distinguish parasite eggs \cite{rodriguez1999detect}. We model the learning process as iterative updates of the individuals' internal templates and quantify the benefits of learning as a function of the learning period. Our results indicate that the first few learning experiences contribute most of the benefits, and therefore imprinting or even an innate template can be more favorable when there are costs associated with learning. We further show that variation among individuals generally makes learning more favorable, whereas variation within clutches will impair the effectiveness of learning. However, in the presence of environmental variation, the latter trend can be reversed. Moreover, environmental variation will cause a reduction in fitness that constitutes an effective cost of learning, which can make learning maladaptive in stochastically changing environments.

\section{Model}

\subsection{Sources of variation}

Consider host birds that lay eggs in clutches of size $N$. We assume that each host egg has a probability $p$ of being replaced by a parasite egg. The hosts may try to distinguish the eggs based on one characteristic, such as color, represented by a trait value $x$. We consider three types of variation among the host eggs. First, there is variation among the eggs of the same individual due to stochasticity in egg production, which will be referred to as ``phenotypic variation'' and will be modeled by a normal distribution $\mathcal{N}(x | \mu_i, \sigma_i^2)$. Second, the mean value $\mu_i$ can vary between individual hosts due to genetic differences, which will be referred to as ``genotypic variation'' and will be modeled by another normal distribution $\mathcal{N}(\mu_i | \mu_h, \sigma_g^2)$; for simplicity, the variance $\sigma_i^2$ is assumed to be the same for all individuals. Thus, the overall distribution of host eggs will be $\mathcal{N}(x | \mu_h, \sigma_h^2)$ with a total variance $\sigma_h^2 = \sigma_g^2 + \sigma_i^2$. We also model all parasite eggs by a normal distribution $\mathcal{N}(x | \mu_p, \sigma_p^2)$; without loss of generality, we assume that $\mu_p > \mu_h$ (Fig.~\ref{fig:illustration}a). Moreover, to study the effect of ``environmental variation'', we assume that the individual mean $\mu_i$ of each host is shifted by a random amount for each clutch. The shifted mean of the $t$-th clutch is modeled by $\mu_t = \mu_i + \eta_t$, where $\eta_t$ is drawn from a normal distribution $\mathcal{N}(\eta_t | 0, \sigma_e^2)$.

\subsection{Rejection threshold and learning}

Following the basic framework of signal detection theory \cite{hautus2021detection} (see Appendix~\ref{sec:SDT}), we assume that hosts use a threshold to decide whether an observed egg is a parasite. Thus, if the observed trait value $x$ of an egg is above a certain threshold $\theta$, it will be rejected as a parasite egg (even if it is the host's own egg). We also assume that, if a parasite egg is not rejected, it will hatch and ruin the whole clutch; otherwise, the fitness of a clutch is given by the number of retained host eggs. Thus, without parasites, the expected fitness of a clutch containing $k$ host eggs when adopting a threshold $\theta$ is $k \, \Phi(\theta|\mu,\sigma^2)$, where $\Phi(x|\mu,\sigma^2)$ is the cumulative probability of a host egg distribution $\mathcal{N}(x|\mu,\sigma^2)$. 
% Note that $\mathcal{N}(x|\mu,\sigma^2)$ is the host's estimate of its own egg distribution, which can be different from the true distribution $\mathcal{N}(x|\mu_i,\sigma_i^2)$. 
With a probability $p$ that each host egg may be replaced by a parasite, the expected fitness of a clutch containing $N$ eggs can be calculated as (see Appendix~\ref{sec:threshold}):
\begin{align}
F(\theta|\mu,\sigma^2) &= N (1-p) \, \Phi(\theta|\mu,\sigma^2) \, \big( 1 - p \, \Phi(\theta|\mu_p,\sigma_p^2) \big)^{N-1} \,,
\end{align}
where $\Phi(x|\mu_p,\sigma_p^2)$ is the cumulative probability of the parasite egg distribution $\mathcal{N}(x | \mu_p, \sigma_p^2)$.

There is a trade-off between having a low threshold that rejects more parasite eggs at the expense of discarding more host eggs, and a high threshold that retains more host eggs but risks accepting parasite eggs. We calculate the optimal threshold by maximizing the expected fitness $F(\theta|\mu,\sigma^2)$ of the clutch. Note that, when the host and parasite distributions overlap, even the optimal threshold can lead to recognition errors. Moreover, because each host individual has a different egg distribution, it is not ideal for all individuals to use the same threshold. Hosts can improve recognition by learning the characteristics of their own eggs to form individual templates. Each individual can observe eggs in its nest to update the template and adjust the threshold. 

\begin{figure}
\centering
\includegraphics[width=\textwidth]{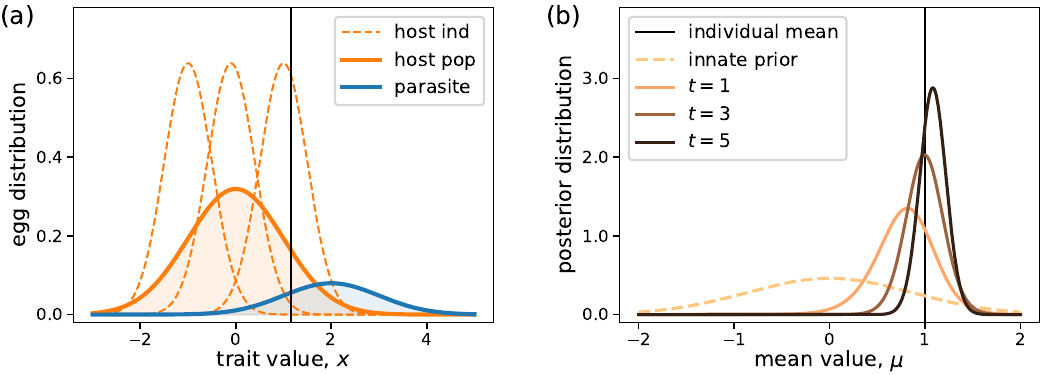}
\caption{\small (a) Egg rejection in the framework of signal detection theory, showing the distribution of parasite eggs (blue curve) weighted by the parasite probability $p$, the egg distribution of the host population (orange solid curve) weighted by $(1-p)$, and the optimal innate threshold for the host population (black vertical line). Individual hosts can have egg distributions with different mean values (orange dashed curves), and would therefore benefit from using thresholds tailored to their individual templates. (b) Updating individual templates through Bayesian learning. All individuals have the same prior distribution of individual mean values before learning (orange dashed curve). The posterior distribution after learning for different numbers of clutches (dark orange curves) moves towards the true individual mean (black vertical line).}
\label{fig:illustration}
\end{figure}

To model how the template is updated after observing new eggs, we follow the idealized scheme of Bayesian learning (Fig.~\ref{fig:illustration}b). Let $P(\mu)$ be the host's prior distribution of its mean $\mu_i$, then after observing an egg $x$, the posterior distribution is given by $P(\mu|x) = P(x|\mu) P(\mu) / \int P(x|\nu) P(\nu) \der\nu$, where the likelihood function is:
\begin{align} \label{eq:likelihood}
P(x|\mu) &= (1-p) \, \mathcal{N}(x | \mu, \sigma_i^2) + p \, \mathcal{N}(x | \mu_p, \sigma_p^2) \,.
\end{align}
The initial prior distribution $P_0(\mu)$ is assumed to be $\mathcal{N}(\mu | \mu_h, \sigma_g^2)$ for all host individuals, which matches their genotypic variation and corresponds to the optimal innate template (see Appendix~\ref{sec:threshold}). To make numerical computation manageable, we approximate the posterior distribution by a normal distribution $\mathcal{N}(\mu|\hat{\mu},\hat{\sigma}^2)$, where $\hat{\mu}$ and $\hat{\sigma}^2$ are the mean and variance of the numerically computed posterior distribution. Here $\hat{\mu}$ represents the host's updated estimate of its true mean $\mu_i$, and $\hat{\sigma}$ represents the uncertainty in the estimate. Taking this uncertainty into account, the host's estimate of its own egg distribution will be $\mathcal{N}(x | \hat{\mu}, \hat{\sigma}^2 \!+\! \sigma_i^2)$. Then, the updated optimal threshold for the host individual will be the $\theta$ value that maximizes the expected fitness $F(\theta | \hat{\mu}, \hat{\sigma}^2 \!+\! \sigma_i^2)$.

To evaluate the fitness benefits of different learning strategies, we simulate a large population of hosts, whose individual mean values $\mu_i$ are randomly drawn from $\mathcal{N}(\mu_i | \mu_h, \sigma_g^2)$. Each host produces a clutch of $N$ eggs at every time step. Each egg's trait value $x$ is drawn from the host individual's egg distribution $\mathcal{N}(x | \mu_i, \sigma_i^2)$; with probability $p$, each egg is replaced with a parasite egg by drawing a new random number $x$ from the parasite's egg distribution $\mathcal{N}(x | \mu_p, \sigma_p^2)$. At each time step, Bayesian learning is performed to update each host's posterior distribution based on every egg in its clutch, then the optimal threshold is calculated numerically and used to reject eggs. After accounting for any parasites that are not rejected, the surviving number of host eggs is recorded and averaged over all individuals to represent the hosts' expected fitness $\bar{F}$.

The model parameters and their default values are listed in Table~\ref{tab:symbols}. For instance, we set $N = 5$ to match the median clutch size of great reed warblers, a well-studied host species in brood parasitism research \cite{rodriguez1999detect}. Note that our parasite probability $p$ is defined per egg, as opposed to the parasitism rate per nest studied in other models \cite{davies1996recognition, rodriguez1999detect}. Our definition allows multiple host eggs to be replaced by parasites within a clutch \cite{Soler2025}, and makes the signal detection and decision making process easier to analyze for each egg. For a clutch size $N = 5$, a probability $p = 0.2$ per egg would imply a 67\% chance that a nest will be parasitized.

\begin{table}[t]
\renewcommand{\arraystretch}{1.15}
\centering
\caption{Variables and parameters in the model} \label{tab:symbols}
\begin{tabular}{cll}
\hline
Symbol & Default Value$^\dag$ & Meaning \\
\hline
$x$ & - & egg trait\\
$\theta$ & - & rejection threshold \\
$T$ & - & learning period (number of clutches) \\
$N$ & 5 & clutch size \\
$p$ & 0.2 & parasite probability (per egg) \\
$\mu_{p}$ & 2.0 & parasite population mean\\	
$\mu_{h}$ & 0 & host population mean \\
$\mu_{i}$ & - & host individual mean \\	
$\sigma_{i}^2$ & 0.25 & host phenotypic variance \\
$\sigma_{g}^2$ & 0.75 & host genotypic variance \\
$\sigma_{e}^2$ & 0 & environmental variance \\
$\sigma_{h}^2$ & 1.0 & host population variance \\
$\sigma_{p}^2$ & 1.0 & parasite population variance \\
$\delta_c$ & - & constitutive cost of learning \\
$\delta_i$ & - & induced cost of learning \\
\hline
\end{tabular} \\[6pt]
$^\dag$The default values are used for generating the figures unless otherwise specified.
\end{table}

\subsection{Learning strategies}

To understand how learning improves the expected fitness, we consider several types of strategies:
\begin{itemize}[noitemsep, topsep=0pt, leftmargin=*]
% \item Perfect decision (PD): This is a hypothetical scenario where the host bird makes the right decision every time -- always accepts its own eggs and rejects parasite eggs.
\item Perfect information (P): This is a hypothetical scenario where the host knows the true distribution of its own eggs as well as that of parasite eggs, and uses the corresponding optimal threshold to reject eggs.
\item Learning (L): The host learns from every clutch of eggs and updates its template. We assume that the host first examines all eggs in the nest to update its threshold, then uses the threshold to accept or reject each egg.
\item Imprinting (M): The host only learns from its first clutch of eggs. After updating once, the template is fixed for the remaining lifetime. We assume that all hosts start with the same initial threshold as in the innate template model.
\item Innate template (I): The host uses a predetermined template; no learning occurs. The rejection threshold is fixed and identical for all individuals. Among all possible thresholds, we consider the one that maximizes the mean fitness of the population.
% \item Acceptor (A): The host bird accepts all eggs in its nest.
\end{itemize}

Among these models, perfect information represents the ideal result of learning, whereas innate template represents the best outcome without learning. Because the innate template does not account for individual variation among hosts, some individuals inevitably commit many false rejections or false acceptances due to unfitting templates. We define the relative fitness gained from learning or imprinting by:
\begin{align}
f = \frac{\bar{F} - \bar{F}_\text{I}}{\bar{F}_\text{P} - \bar{F}_\text{I}} \,,
\end{align}
where $\bar{F}_\text{P}$ is the expected fitness of perfect information and $\bar{F}_\text{I}$ is that of innate template (these fitnesses can be calculated exactly, as shown in Appendix~\ref{sec:threshold}). Thus, the relative fitness of perfect information is normalized to 1, whereas that of innate template is 0.

\section{Results}

\subsection{Benefits and costs of learning}

\begin{figure}[t]
\centering
\includegraphics[width=\textwidth]{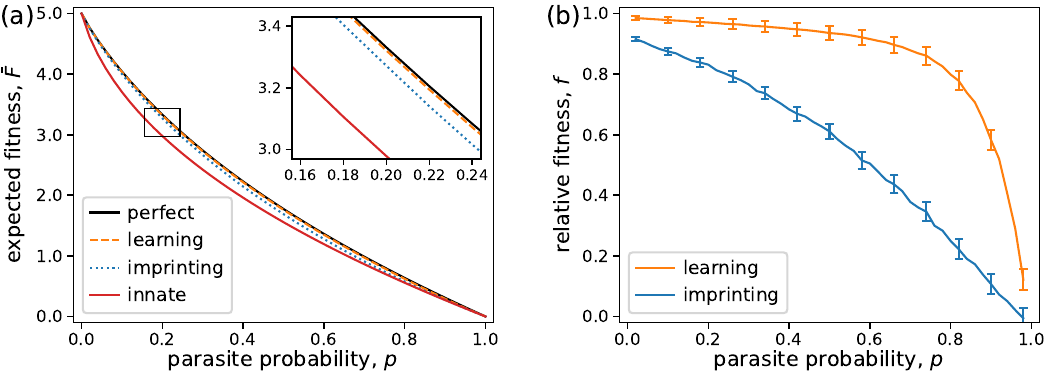}
\caption{\small (a) Expected fitness of different strategies, from high to low: perfect information (black solid curve), learning (orange dashed curve), imprinting (blue dotted curve), and innate template (red solid curve). In all these models, fitness declines with increasing parasite probability. (b) Relative fitness defined by normalizing the expected fitness between those of perfect information ($f=1$) and innate template ($f=0$). The fitness values are calculated from simulations of large populations, and the error bars (shown for every fourth point) represent standard error of the mean.}
\label{fig:models}
\end{figure}

We first study the fitness benefits of learning and imprinting, irrespective of associated costs. In all four models (P, L, M, I), the expected fitness declines with increasing parasitism (Fig.~\ref{fig:models}a), because a higher rate of parasitism reduces the effective number of host eggs contributing to the host's reproductive output. Importantly, for relatively small parasite probability $p$, the relative fitness of learning reaches almost 100\% of perfect information (Fig.~\ref{fig:models}b), meaning that learning substantially improves host defense by adapting rejection thresholds to individual egg distributions. Imprinting, which updates the threshold only once, is able to capture a majority of this fitness benefit (Fig.~\ref{fig:models}b), suggesting that the marginal gains from extended learning is small.

\begin{figure}[t]
\centering
\includegraphics[width=\textwidth]{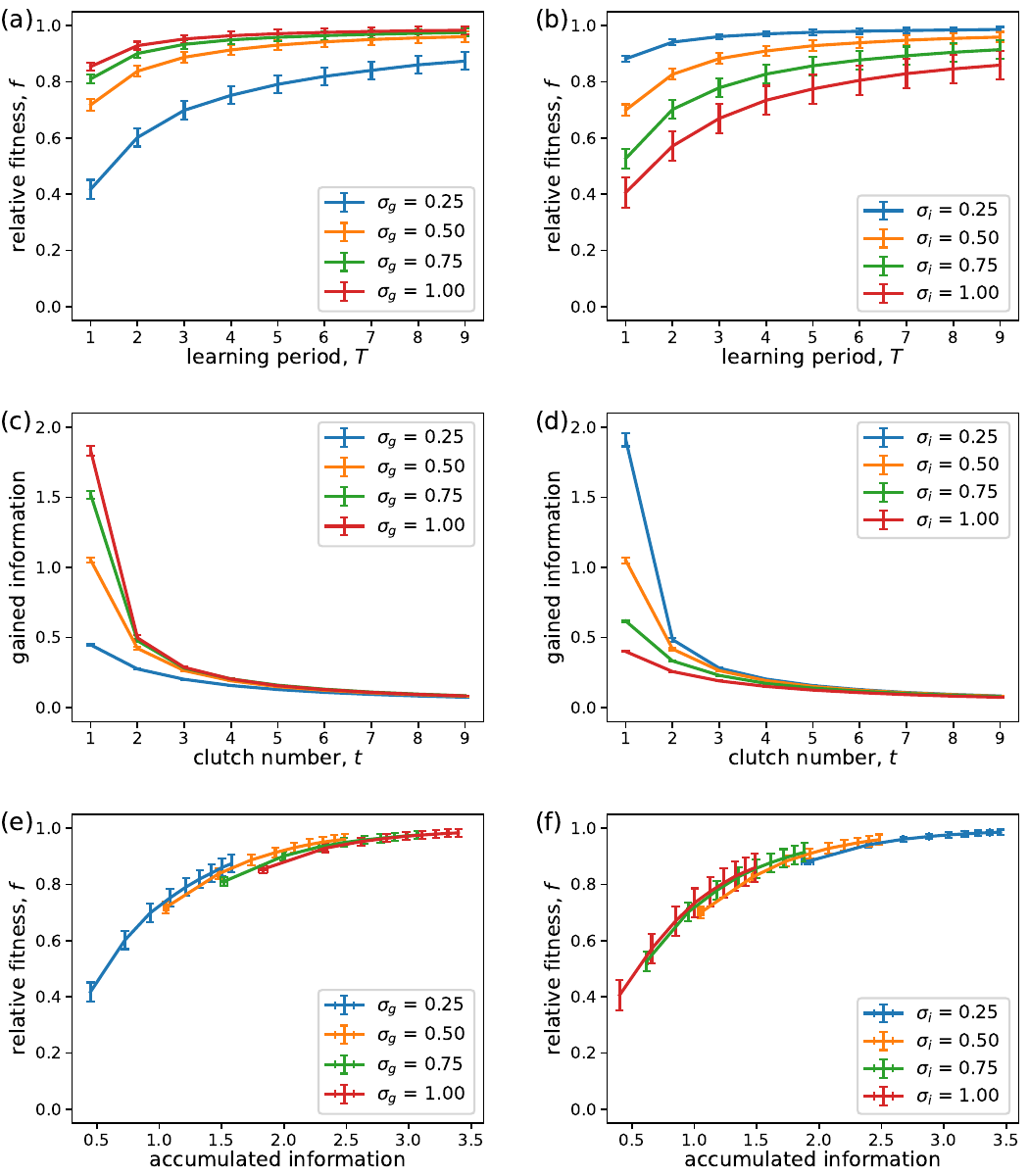}
\caption{\small (a--b) Relative fitness $f$ as a function of the learning period $T$ for different levels of genotypic variation $\sigma_g$ and phenotypic variation $\sigma_i$. Error bars represent standard error of the mean across a simulated population. (c--d) Information about individual mean values gained from observing each clutch, labeled by $t = 1, \cdots, T$. (e--f) Relationship between relative fitness and accumulated information from the entire learning period. Each point along a curve corresponds to a different learning period. Parameters used are: (a,c,e) $\sigma_i = 0.5$; (b,d,f) $\sigma_g = 0.5$; population size: (a,b) 10000; (c,d) 100.}
\label{fig:fit-info}
\end{figure}

The learning and imprinting models differ by the length of the learning period. They represent two extreme cases where the learning period is unlimited or restricted to only one clutch. To study how fitness changes with the learning period, we simulate a population of hosts that learn for $T$ clutches before their rejection thresholds are fixed, and then calculate their expected fitness $\bar{F}$ after those clutches. Thus, $T = 0$ corresponds to the innate template, and $T = 1$ represents imprinting. Fig.~\ref{fig:fit-info}a--b show the relative fitness $f$ as a function of the learning period $T$. There is a large increase in fitness from $T = 0$ ($f = 0$) to $T = 1$, followed by smaller increases at larger $T$, implying that continuing to learn for more clutches does not provide significant gains in fitness.

The diminishing benefits of a long learning period can be understood by calculating the amount of information the hosts gain about their egg appearance from observing each additional clutch. Because the eggs in a clutch are assumed to be independent, the information gained from a clutch is the sum of information from every egg. According to information theory \cite{Cover2005}, the latter can be quantified by the mutual information between an observed egg $x$ and the host's individual mean $\mu$ (see Appendix~\ref{sec:information theory}):
\begin{equation}
I(x; \mu) = \sum_x P(x) \left( \sum_\mu P(\mu|x) \log \frac{P(\mu|x)}{P(\mu)} \right) \,,
\end{equation}
where $P(\mu)$ and $P(\mu|x)$ are the prior and posterior distributions of $\mu$, and $P(x)$ is the marginal distribution of $x$ calculated from the likelihood function (\ref{eq:likelihood}). This mutual information represents how much uncertainty in $\mu$ is reduced by observing $x$ on average. We find that the first clutch provides much more information than later ones. Fig.~\ref{fig:fit-info}c--d show how information learned per clutch decreases with the clutch number. There is a positive relation between the fitness gains and the amount of information acquired (Fig.~\ref{fig:fit-info}e--f), which is reminiscent of similar relations found in other adaptation strategies such as bet-hedging \cite{cohen1967optimizing, Donaldson-Matasci2010, Rivoire2011}. Overall, our results indicate that an extended learning period contributes little to recognition accuracy.

\begin{figure}
\centering
\includegraphics[width=\textwidth]{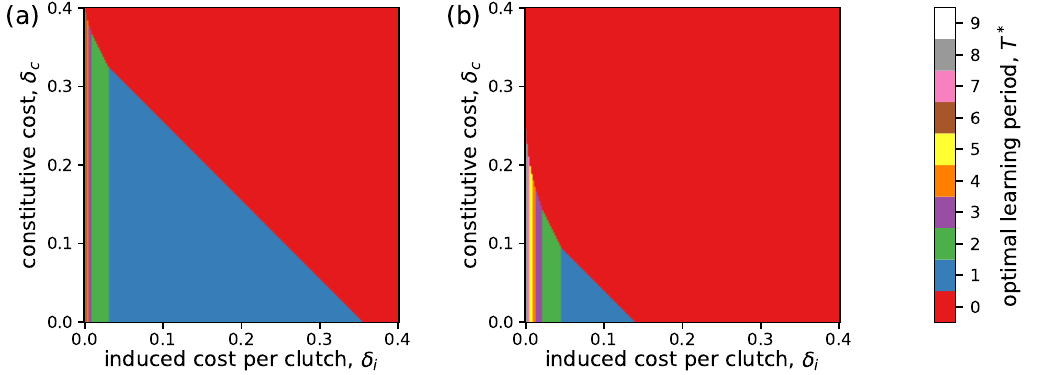}
\caption{\small (a--b) Optimal learning period $T^*$ as a function of the constitutive cost $\delta_c$ and induced cost $\delta_i$ per clutch, in the absence and presence of environmental variation. $T^* = 0$ represents innate template, $T^* = 1$ represents imprinting, and $T^* > 1$ represents learning for multiple clutches. Parameters used are: (a) $\sigma_e = 0$; (b) $\sigma_e = 0.5$; population size: 60000.}
\label{fig:opt_cost}
\end{figure}

The benefits of learning should be weighed against potential costs, including constitutive and induced costs. Since the fitness curves in Fig.~\ref{fig:fit-info}a--b rise sharply, we expect a short learning period to be optimal when there are induced costs associated with extended learning. For simplicity, consider induced costs that are proportional to the learning period, as well as constitutive costs that are one-time investments. The net fitness gain $F_\text{net}(T)$ will be the expected fitness $\bar{F}(T)$ for a given learning period $T$ subtracted by a constitutive cost $\delta_c$ and an induced cost $\delta_i \, T$, where $\delta_i$ is the induced cost per clutch. We calculate the net fitness gain $F_\text{net}(T)$ for different values of $\delta_c$ and $\delta_i$, and numerically search for the optimal learning period $T^*$ that maximizes $F_\text{net}$. Fig.~\ref{fig:opt_cost}a shows how the optimal learning period depends on $\delta_c$ and $\delta_i$. It can be seen that a higher induced cost per clutch leads to a shorter optimal learning period, because of the diminishing benefits of learning. There is a large area in the parameter space where a learning period of $0$ or $1$ is optimal, meaning that either an innate template or imprinting is the most cost-effective strategy.

\subsection{Genotypic and phenotypic variation}

Our goal is to study how the benefits of learning depend on different types of variation in the egg trait, including phenotypic variation among the eggs of the same individual host ($\sigma_i$), genotypic variation of mean egg value among different host individuals ($\sigma_g$), and environmental variation of the mean egg value between different clutches of the same host ($\sigma_e$). We begin by analyzing the effects of genotypic and phenotypic variations in the absence of environmental variation.

\begin{figure}
\centering
\includegraphics[width=\textwidth]{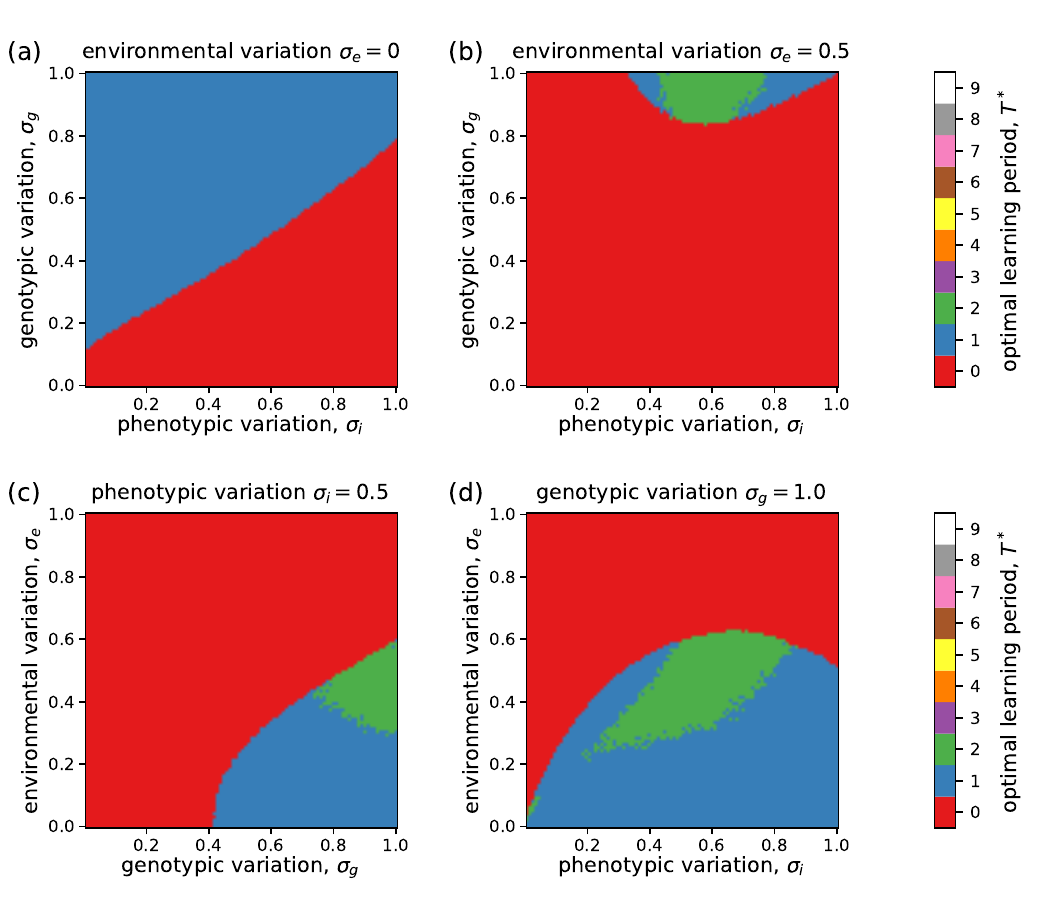}
\caption{\small (a--d) Optimal learning period $T^*$ with respect to genotypic variation $\sigma_g$, phenotypic variation~$\sigma_i$, and environmental variation $\sigma_e$. $T^* = 0$ represents innate template, $T^* = 1$ represents imprinting, and $T^* \!>\! 1$ represents learning for multiple clutches. Parameters used are: $\delta_c = \delta_i = 0.04$; population size: 60000.}
\label{fig:opt_sig}
\end{figure}

As shown in Fig.~\ref{fig:fit-info}a, a larger genotypic variation $\sigma_g$ leads to a faster increase in fitness as a function of the learning period. This can be explained by the greater amount of information gained from each clutch (Fig.~\ref{fig:fit-info}c). When genotypic variation is larger, there is more variation among individuals of the host population, and hence more uncertainty about their egg appearances. Thus, there is more information to be learned, which makes learning more beneficial. Intuitively, when genotypic variation among hosts is large, it pays off to learn a tailored template for oneself.

In contrast, when phenotypic variation $\sigma_i$ increases, the information obtained per clutch decreases (Fig.~\ref{fig:fit-info}d), leading to a lower fitness (Fig.~\ref{fig:fit-info}b). There is less information that hosts can learn from observing eggs when their phenotypic variation is larger. This can be explained by a lower signal-to-noise ratio (SNR) and hence less efficient learning \cite{bialek2012biophysics} (see Appendix~\ref{sec:information theory}). Intuitively, phenotypic variation among eggs in the same clutch acts as noise on top of the individual mean value; the more noise there is, the more difficult it is to estimate the true mean value, hence the template is less accurate.

Fig.~\ref{fig:opt_sig}a shows how the optimal learning period changes with genotypic and phenotypic variations. When genotypic variation increases, the optimal learning period increases because more information can be learned. When phenotypic variation increases, the optimal learning period decreases because learning is inefficient and therefore not cost-effective.

\subsection{Environmental variation}

We next consider environmental variation that can affect egg appearances. We assume that the innate template of the population has adapted to such variation, which means the innate threshold is determined from the stationary distribution of individual means, given by $\mathcal{N} (\mu_i | \mu_h, \sigma_g^2 \!+\! \sigma_e^2)$. However, we assume that individual hosts do not know the statistics of the environmental variation, which means their estimate of $\mu_i$ is still updated using the likelihood function (\ref{eq:likelihood}).

\begin{figure}
\centering
\includegraphics[width=\textwidth]{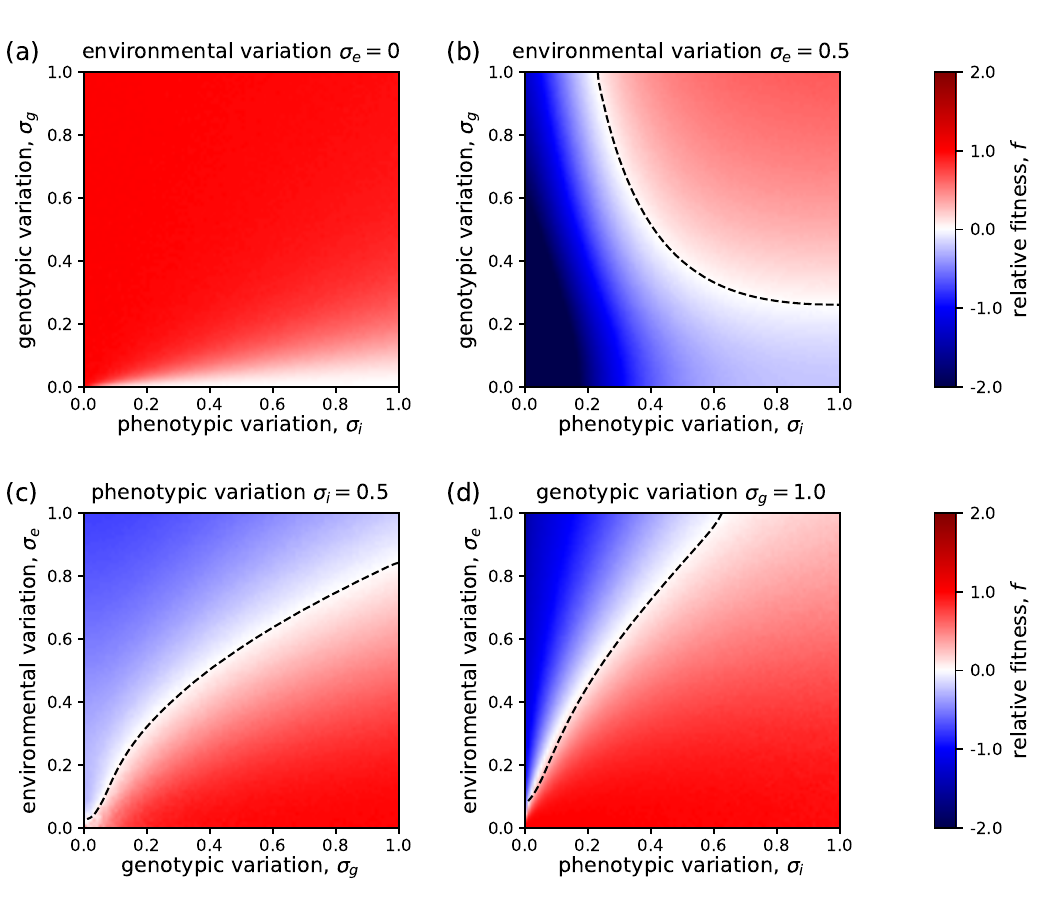}
\caption{\small (a--d) Relative fitness $f$ of learning with respect to genotypic variation $\sigma_g$, phenotypic variation $\sigma_i$, and environmental variation $\sigma_e$, without constitutive and induced costs. (a) In the absence of environmental variation, a positive $f$ (red) means learning is favored. (b--d) Significant environmental variation can make $f$ negative (blue), indicating an effective cost that disfavors learning. Population size for simulations: 60000.}
\label{fig:phase}
\end{figure}

Fig.~\ref{fig:phase}a--b show the relative fitness of learning as a function of genotypic and phenotypic variations in the absence and presence of environmental variation. When environmental variation is absent (Fig.~\ref{fig:phase}a), the relative fitness is always positive, which implies that learning is more advantageous than the innate template. When there is significant environmental variation (Fig.~\ref{fig:phase}b), genotypic variation still enhances the fitness advantage of learning over the innate template; however, phenotypic variation can promote learning rather than inhibit it. Intuitively, even though a large phenotypic variation decreases the accuracy of learning, in a noisy environment the decreased learning accuracy may counterintuitively limit the uptake of misleading information.

Importantly, learning is not always beneficial in varying environments. Besides the parameter region that favors learning (Fig.~\ref{fig:phase}b, red), a new region (blue) now appears, where using an innate template yields higher fitness than learning. These two regions are separated by a ``phase boundary''. Increasing environmental variation will shift this boundary toward the upper-right corner of the diagram. Intuitively, large environmental variation causes the signal to change during the learning period, leading individuals to acquire unreliable information that limits their future fitness.

We can examine how environmental variation interacts with genotypic and phenotypic variations to affect fitness. Fig.~\ref{fig:phase}c shows the relative fitness of learning as a function of the environmental variation $\sigma_e$ and genotypic variation $\sigma_g$. While $\sigma_g$ represents the uncertainty in the target to be learned, $\sigma_e$ represents the unreliability of the signals that individuals are learning from. It can be seen that high signal unreliability and low target uncertainty (Fig.~\ref{fig:phase}c upper-left corner, blue) favor innate template, whereas low signal unreliability and high target uncertainty (lower-right corner, red) promote learning. Such general trends have been discussed in studies of animal behavior \cite{dunlap2009components, dunlap2016reliability}. Fig.~\ref{fig:phase}d shows how the relative fitness of learning depends on the environmental variation $\sigma_e$ and phenotypic variation $\sigma_i$. Consistent with the results above, a large environmental variation makes learning unfavorable compared to the innate template (Fig.~\ref{fig:phase}d upper-left corner, blue). 
% This figure also shows the reversal of the effect of phenotypic variation: when $\sigma_e$ is close to zero, $\sigma_i$ decreases the fitness of learning relative to innate template; but when $\sigma_e$ is sufficiently large, a greater $\sigma_i$ increases the relative fitness of learning.

\subsection{Effective cost of learning}

\begin{figure}
\centering
\includegraphics[width=\textwidth]{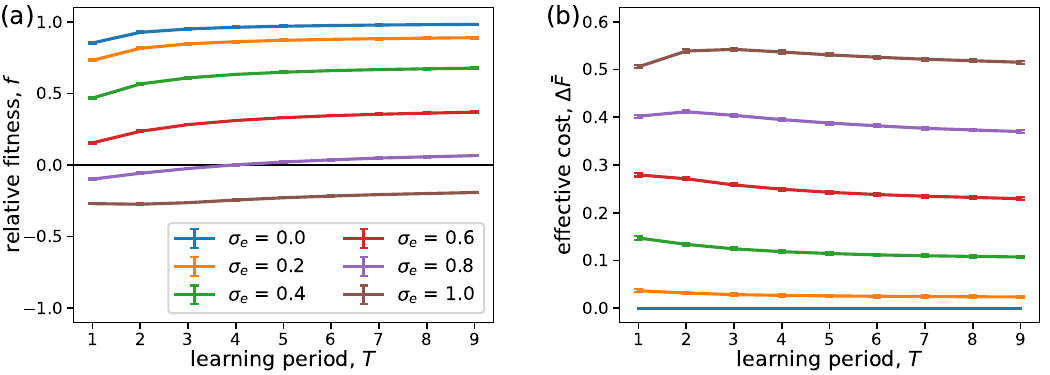}
\caption{\small (a) Relative fitness $f$ of learning as a function of the learning period $T$ under different levels of environmental variation $\sigma_e$, without constitutive and induced costs. (b) Effective cost due to environmental variation, defined as the fitness difference $\Delta \bar{F}$ in the absence and presence of environmental variation. Parameters used are: $\sigma_g = 1.0$, $\sigma_i = 0.5$; population size: 60000.}
\label{fig:env_cost}
\end{figure}

Because environmental variation reduces the fitness of learning, it effectively constitutes a cost. To characterize this effective cost of learning, we study how it depends on the learning period $T$. Fig.~\ref{fig:env_cost}a shows the relative fitness of learning as a function of the learning period for different levels of environmental variation $\sigma_e$. It can be seen that, for a sufficiently large $\sigma_e$ (e.g., around $0.8$), the curve starts below zero at $T = 1$, implying that imprinting becomes less favorable than the innate template. For even larger $\sigma_e$, the entire curve falls below zero, meaning that extended learning is also disfavored, in agreement with Fig.~\ref{fig:phase}c--d (blue).

We can define the effective cost of learning due to environmental variation by the difference in the expected fitness $\bar{F}$ between $\sigma_e = 0$ (no environmental variation) and a nonzero $\sigma_e$. Fig.~\ref{fig:env_cost}b shows such effective cost as a function of the learning period $T$. Recall that a constitutive cost would be a constant value independent of $T$, whereas an induced cost is expected to increase with $T$. It can be seen that the effective cost due to environmental variation remains roughly constant, and thus behaves largely like a constitutive cost. This is understandable because environmental variation is always present and causes the individual means to fluctuate around their true values; such fluctuation does not increase with the learning period and does not vanish after individuals stop learning. Learning is favored only when the benefits of learning outweigh the effective cost.

Fig.~\ref{fig:opt_cost}b shows how the optimal learning period $T^*$ changes in the presence of environmental variation. Because its effective cost largely adds onto the constitutive cost, the phase boundaries are moved downward and the $T^* > 0$ phases are squeezed into the lower-left corner of the parameter space. In particular, a wide range of constitutive and induced costs that support imprinting in a stable environment now favor an innate template instead. 

Fig.~\ref{fig:opt_sig}c shows how the optimal learning period changes with the genotypic variation $\sigma_g$ and environmental variation $\sigma_e$ for a given constitutive cost $\delta_c$ and induced cost $\delta_i$. Similarly, Fig.~\ref{fig:opt_sig}d shows the effect of the phenotypic variation $\sigma_i$ and environmental variation $\sigma_e$. In both figures, it can be seen that increasing environmental variation eventually makes the innate template favorable. Interestingly, however, the optimal learning period can go from $T^* = 1$ (imprinting) to $T^* > 1$ before jumping to $T^* = 0$ (innate template). This would not be possible if the effective cost from environmental variation behaves strictly as a constitutive cost. Indeed, it can be seen from Fig.~\ref{fig:env_cost}b that, for some $\sigma_e$, the effective cost slightly goes down with the learning period $T$, which is why a longer learning period $T > 1$ can be favored over $T = 1$.

\subsection{Egg mimicry, clutch size, and lifespan}

\begin{figure}[h!]
\centering
\includegraphics[width=\textwidth]{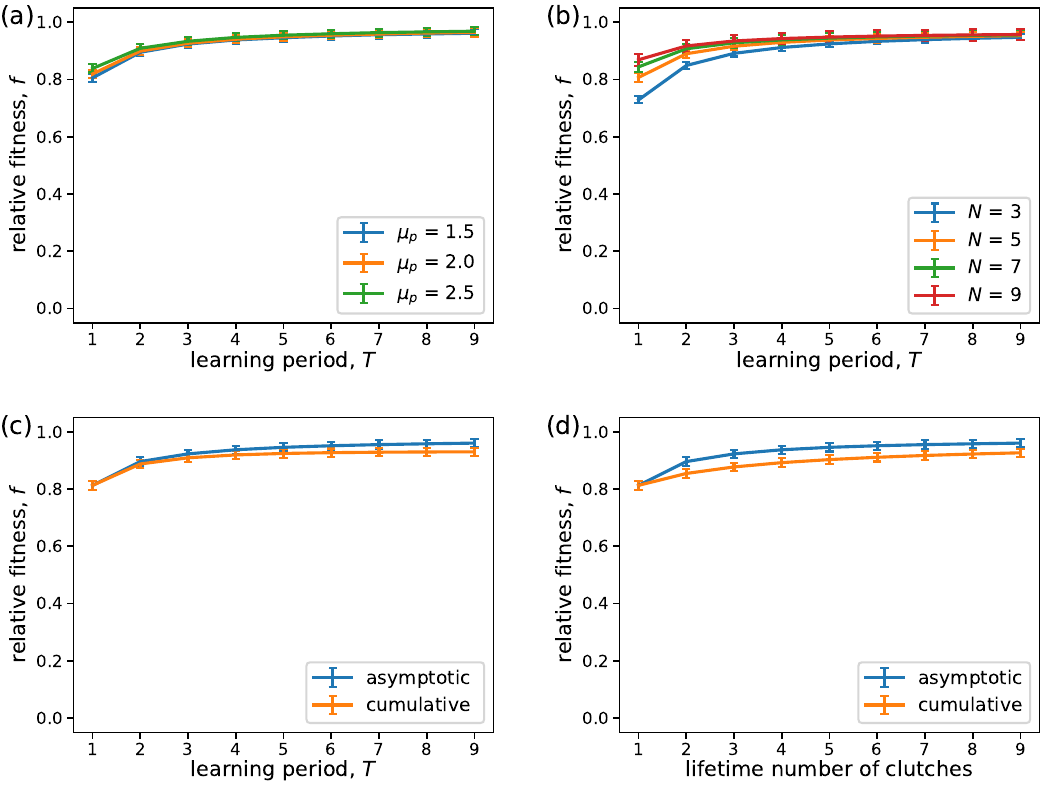}
\caption{\small (a) Relative fitness $f$ as a function of the learning period $T$ for different parasite mean values $\mu_p$. The near collapse of the curves suggests that $\mu_p$, or equivalently the signal discriminability $d' = \mu_p - \mu_h$, has little effect on the relative fitness. (b) $f$ versus $T$ for different clutch sizes $N$. The curves show the same trend of sharp increase at small $T$ and leveling off at large $T$. (c) Relative fitness calculated from two fitness measures: asymptotic fitness (blue) defined as the expected fitness after learning is complete, and cumulative fitness (orange) represented by the average fitness per clutch over a host's lifetime. Both fitnesses rise sharply with the learning period $T$ and approach similar values. (d) Asymptotic (blue) and cumulative (orange) fitness versus the total number of clutches under extended learning throughout the lifespan. Parameters used are: population size: 10000; (c) lifetime number of clutches: 10.}
\label{fig:discussion}
\end{figure}

To verify that our results are robust with respect to other parameters of the model, we have systematically varied the parameter values, including the mean value of the parasite egg distribution ($\mu_p$), the clutch size ($N$), and the lifespan of the host birds. First, since we have chosen the overall egg distribution of the host population to be centered at $\mu_h = 0$, the parameter $\mu_p$ represents the distance $d'$ between the means of the host and parasite egg distributions. In signal detection theory, $d'$ is known as the discriminability of the signal \cite{hautus2021detection}. The smaller $d'$ is, the more overlap there is between the distributions, and the more difficult it is for the host to recognize parasite eggs (See Appendix~\ref{sec:SDT}). Some brood-parasitic birds have evolved to produce eggs that mimic key traits of their hosts' eggs, making them nearly indistinguishable in color, pattern, shape, or size \cite{dillenseger2024active, stevens2013bird, stoddard2011avian}. In our model, closer resemblance can be represented by a smaller $d'$, which will lower the hosts' fitness $\bar{F}$ due to reduced discriminability. However, the relative fitness $f$ is largely unaffected by $d'$ (Fig.~\ref{fig:discussion}a). Thus, by focusing on the relative fitness, our results are insensitive to the value of $\mu_p$.

We have assumed, for simplicity, that all clutches have the same size $N = 5$. A larger clutch size can provide the host with a larger sample for learning, but the cost of missing a parasite egg is also greater. We evaluated the effect of clutch size by varying $N$ while holding the parasite probability $p$ constant. Fig.~\ref{fig:discussion}b shows that clutch size has only a minor effect on the benefits of learning, which does not change the sharp increase of relative fitness with learning period. This indicates that our results remain robust for different clutch sizes.

Throughout our model, we have measured fitness by the expected number of surviving host eggs per clutch after the learning period is complete. This represents the ``asymptotic'' outcome of learning. An alternative measure of fitness is the average number of surviving host eggs over all clutches during a host's lifetime, which represents the ``cumulative'' benefits of learning. As shown in Fig.~\ref{fig:discussion}c, the cumulative fitness increases with the learning period in a pattern similar to that of the asymptotic fitness. If the host lifespan is long, such as magpies that can live for over 5 years \cite{martinez2020age}, the difference between the cumulative and asymptotic fitness is insignificant. However, some host birds have relatively short lifespans; for example, the average lifespan of a female great reed warbler in the wild is around $2.6$ years \cite{hansson2004lifetime}. In such cases, the cumulative fitness may be a more meaningful measure. Nevertheless, Fig.~\ref{fig:discussion}d shows that cumulative fitness is still close to the asymptotic fitness even when the lifespan is short.

\section{Discussion}

\subsection{Predictions of the model}

By weighing the benefits and costs of learning, theoretical models have provided arguments as to when different learning strategies, such as imprinting or innate template, are most favorable \cite{shizuka2020learn, rodriguez1999detect}. However, the parameters of these models, such as the cost of learning or the rate of perceptive error, are often hard to measure in experiments, making it difficult to test model predictions. We have chosen to focus on the effects of phenotypic, genotypic, and environmental variations in the learning signal, which are potentially measurable in the field \cite{dunlap2009components, dunlap2016reliability}. For brood parasitism, these correspond to the egg trait variation within a clutch of the same individual ($\sigma_i$), variation between different individuals ($\sigma_g$), and variation among different clutches over time ($\sigma_e$). These components of variation can be evaluated in field experiments \cite{Lotem_1995, rodriguez1999detect}. Our model makes general predictions about the effect of these variations. For example, the same parasite species may exploit multiple host species with different levels of variation in egg appearance; even within a host species, local populations in separate geographic regions can exhibit different levels of variation \cite{yang2014egg, soler1999genetic}. Our model predicts that higher genetic variation within a host population would favor the evolution of learning.

Our model also predicts that environmentally induced variation in egg appearance can influence the selection on egg rejection strategies. When a species’ egg appearance is affected by environmental conditions, fluctuations in the environment will result in an additional cost that can offset the benefits of learning. For example, in great reed warblers, eggs laid under colder conditions tend to be darker, indicating that eggshell pigmentation can shift in response to short-term environmental variation \cite{honza2012weather}. Our model predicts that host populations living in more rapidly changing climates would rely more on an innate template rather than learning. This prediction can be tested by comparing host populations that experience different levels of environmental variability.

\subsection{Experimental studies of learning and imprinting}

\begin{figure}[t]
\centering
\includegraphics[width=\textwidth]{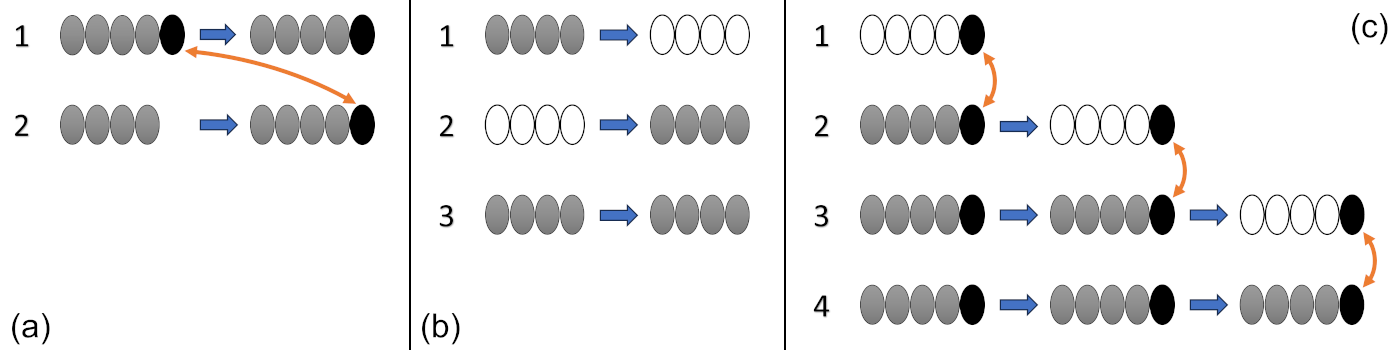}
\caption{\small (a) Illustration of the re-nesting experiment by Lotem et al. \cite{Lotem_1995}. Each egg in a nest was replaced with a lightly painted (gray) or darkly painted (black) egg shortly after it was laid. Learning was evaluated by comparing rejection rates of the darkly painted egg in the first clutch of group 1 and the second clutch of group 2 (orange arrows). (b) Illustration of the imprinting experiment by Soler et al. \cite{Soler_2013}. Each clutch was either manipulated (gray) or unmanipulated (white); eggs in a manipulated clutch were painted as soon as they were laid. Imprinting would imply rejection of eggs in the second clutch that differed in appearance from the first clutch, as in group 1 and 2. (c) Proposed experimental design that can be used to test the length of the learning period.}
\label{fig:experiment}
\end{figure}

The learning mechanisms of host birds have previously been studied in experiments. Notably, Lotem et al. \cite{Lotem_1995} found that host birds can learn from eggs in their first nest but tend to stop learning as they age. In a field experiment with the great reed warbler, the host eggs were replaced with lightly painted eggs and a darkly painted test egg after the full clutch was laid (illustrated as group 1 in Fig.~\ref{fig:experiment}a); the rejection rate of the test egg was recorded. The breeding time was used as a proxy for host age, because more experienced host birds tend to breed early in the season. It was found that naive breeders had a significantly lower rejection rate than experienced breeders. Furthermore, re-nesting experiments were conducted by removing all eggs from a host nest to induce the laying of a second clutch, from which one egg was replaced with a darkly painted egg (group 2 in Fig.~\ref{fig:experiment}a). It was found that older females had almost the same rejection rates between the first and second clutches, whereas younger females showed a higher rejection rate after having the chance to observe their first clutch. However, the experiments had some limitations. Because they were conducted in the wild, it was difficult to know exactly how many breeding seasons a host bird had before the study \cite{Lotem_1995}. In addition, the rejection rate of group 1's first clutch was compared with that of group 2's second clutch (orange arrows in Fig.~\ref{fig:experiment}a), which has the caveat that the first group had observed only one clutch of eggs, whereas the second group had observed two.

Soler et al. \cite{Soler_2013} conducted experiments to test imprinting behavior using captive house sparrows that were first-time breeders and laid multiple clutches. Three experimental groups were studied, one with a manipulated first clutch and unmanipulated second clutch, one with two clutches in the opposite order, and one with both clutches manipulated (illustrated as group 1--3 in Fig.~\ref{fig:experiment}b). All eggs in a manipulated clutch were painted with dark spots as soon as they were laid; rejection rates of the second clutch were evaluated to determine whether hosts rejected eggs that differed in appearance from their first clutch. In principle, the effect of the first clutch on learning can be evaluated by comparing rejection rates between group 2 and 3, and the effect of the second clutch can be evaluated by comparing group 1 and 3. However, no egg rejection was observed in any of these groups, potentially because the manipulation was not strong enough to provoke rejection \cite{Soler_2013}. %In addition, their experimental design would not have allowed studying whether learning continues in the second clutch.

Based on these previous experiments, we propose a modified experimental design to test learning and imprinting. To control for breeding history, it is better to use captive first-time breeders. Our proposed egg manipulations of different groups are illustrated in Fig.~\ref{fig:experiment}c, where the gray eggs represent lightly painted eggs that are used to replace the host eggs immediately after they are laid, and the last egg of a clutch is replaced with a darkly painted egg as the test egg. By comparing the rejection rates of test eggs in the first clutch of group 1 and 2, we can evaluate whether hosts learn from the first clutch, as host birds often take a few days to observe and decide whether to reject certain eggs \cite{Lotem_1995}. Similarly, the influence of the second clutch can be evaluated by comparing the rejection rates in the second clutch between group 2 and 3. The advantage of our experimental design is that hosts from group 2 and 3 share the same history of clutch manipulation except for the second clutch, which allows clear determination of its influence on learning. Likewise, the influence of the third clutch can be evaluated by comparing group 3 and 4, which share the same manipulation history except for the third clutch, and so on. This experimental design makes it possible to determine whether host birds learn from each clutch and when they cease to learn.

\section{Conclusion}

Learning enables hosts to improve parasite recognition by updating their internal templates based on observation of their own clutches. Compared with an innate template shared among all individuals, learning can substantially increase the expected fitness of the host by creating a tailored template for each individual. We have shown that most of this improvement occurs in the early learning period, such that extended learning after the first few clutches does not significantly enhance fitness. The trade-off between benefits and costs of learning leads to different optimal strategies, depending on various sources of variation in egg appearances. In particular, with induced costs that increase with the learning period, the optimal strategy often favors imprinting or even an innate template. Moreover, we have found that environmental variation distorts the signal used for learning and creates an effective cost that can make learning maladaptive.

Learning is typically understood as the process of organisms using external signals to infer environmental states. However, egg recognition provides an interesting case of hosts learning about their own intrinsic properties (i.e., their egg appearances). This form of ``introspective'' learning is similar to other biological systems in which organisms adapt through acquiring information about their own kind. For example, in quorum sensing \cite{coolahan2025review}, bacterial populations monitor their own density to regulate gene expression, and in our body, the adaptive immune system learns to recognize pathogens from self antigens through prior exposures \cite{natoli2019adaptation, netea2019innate}. Such cases highlight that learning is not restricted to external environments, broadening our understanding of how the adaptive use of information evolves across biological systems.

In our model, we considered a one-dimensional signal for simplicity. In nature, however, host birds may rely on multiple egg traits simultaneously, integrating cues such as color, pattern, and size \cite{fulmer2022review, spottiswoode2010visual}. Moreover, the decision rules governing egg rejection are likely to be far more complex than a simple threshold applied to a single trait. Extending the model to a multidimensional recognition framework \cite{kikuchi2015empirical}, such as through clustering or classification approaches, will be an important direction for future work. In addition to egg recognition, many host species also use hatchling recognition as a defense against brood parasites \cite{shizuka2020learn, lotem1993learning}. Incorporating multiple recognition stages in the model will help reveal how hosts coordinate different cognitive strategies and whether learning at one stage influences decisions at another.

Imprinting is not unique to egg recognition; it is a widespread learning mechanism across biological systems. A short but highly informative learning period can have lifelong consequences for behavior and fitness, as in the cases of newly hatched ducklings that follow the first moving objects they see to obtain parental care \cite{lorenz1937companion}, and juvenile salmon that memorize olfactory cues of their natal streams to guide future return from the ocean for reproduction \cite{dittman1996homing}. Moreover, at the microscopic level, various mechanisms of epigenetic inheritance allow a different kind of imprinting in both macro- and microorganisms, which provides a phenotypic memory that facilitates adaptation in fluctuating environments \cite{bonduriansky2009nongenetic, Heard2014transgenerational}. A trade-off between the benefits and effective costs of learning, as studied in our model, may potentially explain the evolution of imprinting under those circumstances. We hope that our theoretical framework will offer broader insights into learning and imprinting as complex and fascinating adaptation strategies in nature.

\appendix

\section{Signal Detection Theory} \label{sec:SDT}

Signal Detection Theory (SDT) provides a way to describe and analyze decision making when the process contains uncertainty \cite{green1966signal, hautus2021detection}. Originally developed for human perception and communication systems, SDT has since become widely used in behavioral ecology, including studies of avian brood parasitism, where hosts must distinguish their own eggs from those of a parasitic species \cite{davies1996recognition, Suarez2020signal, Ruiz-Raya2020signal}.

In the simplest case, each stimulus is represented by a one-dimensional value $x$ along a continuous axis (e.g., one characteristic of egg appearances). The response is determined by a threshold $\theta$ along this axis: without loss of generality, values above the threshold will lead to rejection (as parasite eggs), and values below are accepted (as host eggs). Two types of errors can occur: ``false alarms'' (false positives), in which a host egg is rejected as if it were a parasite egg, and ``misses'' (false negatives), in which a parasite egg is mistakenly accepted as a host egg. The fitness consequences of these errors are often asymmetric. For example, accepting a parasite egg can ruin a whole clutch, whereas rejecting a host egg only reduces a portion of the host’s reproductive output.

SDT characterizes responses using the false alarm (false positive) rate $\alpha = P(x \!>\! \theta | \text{host})$ and the ``hit'' (true positive) rate $\gamma = P(x \!>\! \theta | \text{parasite})$. By shifting the threshold $\theta$, one obtains a family of $(\alpha, \gamma)$ pairs that form the receiver operating characteristic (ROC) curve. The shape of the ROC curve depends on how separable the two signal distributions are. The separation of their means (relative to their variances) is defined as the discriminability index $d'$. A larger $d'$ indicates better discrimination, producing ROC curves that approach the ideal point $(0,1)$, where false alarms are eliminated and all hits are detected.

\section{Calculation of Fitness and Optimal Threshold} \label{sec:threshold}

Let $\mathcal{N}(x | \mu, \sigma^2)$ be the distribution of host eggs, with cumulative probability distribution $\Phi(x | \mu, \sigma^2)$, and $\mathcal{N}(x | \mu_p, \sigma_p^2)$ be the distribution of parasite eggs, with cumulative probability distribution $\Phi(x | \mu_p, \sigma_p^2)$. For a threshold $\theta$, the expected fitness of a clutch containing $k$ host eggs and no parasite eggs is $k \, \Phi(\theta|\mu,\sigma^2)$. For a clutch containing $N$ eggs, each with a probability $p$ of being a parasite, the expected fitness can be calculated as
\begin{align}
F(\theta|\mu,\sigma^2) &= \sum^{N}_{k=0} \binom{N}{k} (1-p)^k \, p^{N-k} \big( 1 - \Phi(\theta|\mu_p,\sigma_p^2) \big)^{N-k} \cdot k \, \Phi(\theta|\mu,\sigma^2) \nonumber \\
&= N (1-p) \, \Phi(\theta|\mu,\sigma^2) \, \big( 1 - p \, \Phi(\theta|\mu_p,\sigma_p^2) \big)^{N-1} \,,
\end{align}
where we used the mathematical identity $\sum^{N}_{k=0} \binom{N}{k} \, k \, a^k \, b^{N-k} = N \, a \, (a+b)^{N-1}$.

For a host with perfect information of its individual mean $\mu_i$ and variance $\sigma_i^2$, the expected fitness is $F(\theta | \mu_i, \sigma_i^2)$. Let $\theta_i (\mu_i)$ be the optimal threshold that maximizes the fitness $F(\theta | \mu_i, \sigma_i^2)$ for the individual, then the expected fitness for the host population is
\begin{equation}
\bar{F}_\textrm{P} = \int_{-\infty}^{\infty} F(\theta_i (\mu_i) | \mu_i, \sigma_i^2) \, \mathcal{N}(\mu_i | \mu_h, \sigma_h^2) \der\mu_i \,.
\end{equation}
We calculate this value numerically using SciPy's \texttt{optimize} and \texttt{integrate} modules.

For the innate template model, the optimal threshold $\theta_0$ for all individuals is the threshold that maximizes the population mean fitness
\begin{align}
\bar{F}(\theta) &= \int_{-\infty}^{\infty} F(\theta | \mu_i, \sigma_i^2) \, \mathcal{N}(\mu_i | \mu_h, \sigma_g^2) \der\mu_i \nonumber \\
&= N (1-p) \big( 1 - p \, \Phi(\theta | \mu_p, \sigma_p^2) \big)^{N-1}  \int_{-\infty}^{\infty} \Phi(\theta | \mu_i, \sigma_i^2) \, \mathcal{N}(\mu_i | \mu_h, \sigma_g^2) \der\mu_i \,.
\end{align}
Using the mathematical identity $\int_{-\infty}^{\infty} \Phi(a|x,b^2) \, \mathcal{N}(x|\mu,\sigma^2) \der x = \Phi(a|\mu,b^2 \!+\!  \sigma^2)$, we can simplify this to
\begin{align}
\bar{F}(\theta) = N (1-p) \big( 1 - p \, \Phi(\theta | \mu_p, \sigma_p^2) \big)^{N-1} \, \Phi(\theta | \mu_h, \sigma_h^2) \,,
\end{align}
where $\sigma_h^2 = \sigma_i^2 + \sigma_g^2$. This expression exactly matches that of $F(\theta | \mu_h, \sigma_h^2)$. Let this be maximized by a threshold $\theta_0$, then the expected fitness for the innate template model is $\bar{F}_\textrm{I} = F(\theta_0 | \mu_h, \sigma_h^2)$. This value is calculated numerically using SciPy's \texttt{optimize} module.

Note that $F(\theta | \mu_h, \sigma_h^2)$ is also the expected fitness of a host whose uncertainty of its mean $\mu$ is given by the distribution $\mathcal{N}(\mu | \mu_h, \sigma_g^2)$, i.e., the initial prior distribution for learning individuals. Therefore, $\theta_0$ is the same as the initial threshold for all individuals in the learning model. However, the threshold used by individuals to reject eggs in even the first clutch will be different from this innate threshold, because we assume that individuals observe and learn from all eggs in a clutch before deciding to reject any of them.

\section{Information Theory}
\label{sec:information theory}

Learning helps reduce uncertainty about unknown variables. Such uncertainty can be quantitatively measured by the entropy of the variable \cite{Cover2005}. A system with high entropy has many possible states; through repeated observations, organisms accumulate information that constrains these possibilities. In this sense, learning can be viewed as a process of entropy reduction \cite{bialek2012biophysics}.

The information gained by the host from learning can be calculated as follows. For a prior distribution $P(\mu)$ of the individual mean $\mu$, the entropy is given by \cite{Cover2005}
\begin{equation}
H[P(\mu)] = - \sum_\mu P(\mu) \log P(\mu) .
\end{equation}
After observing a data point $x$, the entropy is reduced to
\begin{equation}
H[P(\mu|x)] = - \sum_\mu P(\mu|x) \log P(\mu|x) ,
\end{equation}
where $P(\mu|x)$ is the posterior distribution of $\mu$. Thus, on average, the amount of entropy reduction is given by
\begin{align}
I(\mu; x) &= \sum_x P(x) \Big( H[P(\mu)] - H[P(\mu|x)] \Big) \nonumber \\
&= \sum_x \sum_\mu P(x) P(\mu|x) \log \frac{P(\mu|x)}{P(\mu)} \,,
\end{align}
where $P(x)$ can be calculated as $P(x) = \sum_\nu P(x|\nu) P(\nu)$.
This quantity $I(\mu;x)$ is known as the mutual information, and is symmetric in $\mu$ and $x$.

The mutual information is closely related to the signal-to-noise ratio (SNR), which can be defined as the ratio of the variance in the signal to the variance of the effective noise \cite{bialek2012biophysics}. According to the Shannon–Hartley theorem \cite{Cover2005}, the SNR determines the maximum amount of information that can be transmitted reliably. In our model, the SNR can be expressed as $\sigma_g^2 / \sigma_i^2$, which suggests that a higher $\sigma_g$ or a lower $\sigma_i$ would lead to more information gain and hence more effective learning.

\section*{Acknowledgements}
We would like to thank Mathew Leibold and other participants of the PErspectives in Ecological ReSearch (PEERS) seminar at UF for helpful comments on our work.

\printbibliography

\end{document}